# Collapse and Revival of the Matter Wave Field of a Bose-Einstein Condensate

Markus Greiner, Olaf Mandel, Theodor W. Hänsch & Immanuel Bloch[*]

Sektion Physik, Ludwig-Maximilians-Universität, Schellingstrasse 4/III, D-80799 Munich, Germany & Max-Planck-Institut für Quantenoptik, D-85748 Garching, Germany.

**At the heart of a Bose-Einstein condensate lies its description as a single giant matter wave. Such a Bose-Einstein condensate represents the most "classical" form of a matter wave, just as an optical laser emits the most classical form of an electromagnetic wave. Beneath this giant matter wave, however, the discrete atoms represent a crucial granularity, i.e. a quantization of this matter wave field. Here we show experimentally that this quantization together with the cold collisions between atoms lead to a series of collapses and revivals of the coherent matter wave field of a Bose-Einstein condensate. We observe such collapses and revivals directly in the dynamical evolution of a multiple matter wave interference pattern, and thereby demonstrate a striking new behaviour of macroscopic quantum matter.**

While the matter wave field of a Bose-Einstein condensate is usually assumed to be intrinsically stable, apart from incoherent loss processes, it has been pointed out that this should not be true when a Bose-Einstein condensate is in a coherent superposition of different atom number states[1-6]. This is the case for example whenever a Bose-Einstein condensate is split into two parts, such that a well-defined relative phase between the two matter wave fields is established. When these two condensates are isolated from each other, a fundamental question arises: How does the relative phase between the two condensates evolve and what happens to the individual matter wave fields? In this work, we show that in such a case the matter wave field of a Bose-Einstein condensate undergoes a periodic series of collapses and revivals due to the quantized structure of the matter wave field and the collisions between individual atoms.

In order to determine the time evolution of a many atom state with repulsive interactions in a confining potential, let us first assume that all atoms occupy only the ground state of the external potential. The Hamiltonian governing the system after subtracting the ground state energy of the external potential is then solely determined by the interaction energy between the atoms:

$$\hat{H} = \frac{1}{2} U \hat{n}(\hat{n} - 1). \qquad (1)$$

Here $\hat{n}$ counts the number of atoms in the confining potential and $U$ is the on-site interaction matrix element that characterizes the energy cost due to the repulsive interactions when a second atom is added to the potential well. It can be related to the $s$-wave scattering length $a$ and the ground state wave function $w(\mathbf{x})$ through $U = 4\pi\hbar^2 a / m \int |w(\mathbf{x})|^4 d^3x$, as long as the vibrational level spacing of the external potential is large compared to the interaction energy.

The eigenstates of the above Hamiltonian are Fock states $|n\rangle$ in the atom number, with eigenenergies $E_n = U n(n-1)/2$. The time evolution of such an $n$-particle state is then simply given by $|n\rangle(t) = |n\rangle(0) \times \exp(-iE_n t/\hbar)$.

Let us now consider a coherent state $|\alpha\rangle$ (see e.g. ref. 7) of the atomic matter field in a potential well. Such a coherent state with a complex amplitude $\alpha$ and an average number of atoms $\bar{n} = |\alpha|^2$ can be expressed as a superposition of different number states $|n\rangle$ such that $|\alpha\rangle = \exp(-|\alpha|^2/2) \sum_n \frac{\alpha^n}{\sqrt{n!}} |n\rangle$. Now the system is in a superposition of different eigenstates, which evolve in time according to their eigenenergies $E_n$. This allows us to calculate the time evolution of an initially coherent state

$$|\alpha\rangle(t) = e^{-|\alpha|^2/2} \sum_n \frac{\alpha^n}{\sqrt{n!}} e^{-i\frac{1}{2}U n(n-1)t/\hbar} |n\rangle. \qquad (2)$$

Evaluating the atomic field operator $\hat{a}$ for such a state then yields the macroscopic matter wave field $\psi = \langle \alpha(t)|\hat{a}|\alpha(t)\rangle$ with an intriguing dynamical evolution. At first, the different phase evolutions of the atom number states lead to a collapse of the macroscopic matter wave field $\psi$. However, at times $t=l \cdot t_{rev} = l \cdot h/U$, with $l$ being an integer, all phase factors in the sum of equation (2) re-phase to integer multiples of $2\pi$, thus leading to a perfect revival of the initial coherent state. The collapse time $t_c$ depends on the variance $\sigma_n^2$ of the atom number distribution, such that $t_c \approx t_{rev}/\sigma_n$ (see refs. 1-5). A more detailed picture of the dynamical evolution of the macroscopic matter wave field can be seen in Figure 1, where the overlap of an arbitrary coherent state $|\beta\rangle$ with the state $|\alpha\rangle(t)$ is shown for different evolution times up to the first revival time of the many body state[8,9].

In our experiment, we create coherent states of the matter wave field in a potential well, by loading a magnetically trapped Bose-Einstein condensate into a three-dimensional optical lattice potential. For low potential depths, where the tunnelling energy $J$ is much larger than the onsite repulsive interaction energy $U$ in a single well, each atom is spread out over all lattice sites. For the case of a homogeneous system with $N$ atoms and $M$ lattice sites the many body state can then be written in second quantization as a product of identical single particle Bloch waves with zero quasi-momentum

$|\Psi_{SF}\rangle_{U/J=0} \propto \left(\sum_{i=1}^{M} \hat{a}_i^\dagger\right)^N |0\rangle \approx \prod_{i=1}^{M} |\phi_i\rangle$. In the limit of large $N$ and $M$ the atom number distribution of $|\phi_i\rangle$ in each potential well is Poissonian and almost identical to that of a coherent state. Furthermore, all the matter waves in different potential wells are phase coherent, with constant relative phases between lattice sites. As the lattice potential depth $V_A$ is increased and the tunnel coupling $J$ decreases, the atom number distribution in each potential well becomes pronounced sub-Poissonian[10] due to the repulsive interactions between the atoms, even before entering the Mott insulating state[11-13]. After preparing superposition states $|\phi_i\rangle$ in each potential well, we increase the lattice potential depth rapidly in order to create isolated potential wells. The Hamiltonian of eq. 1 then determines the dynamical evolution of each of these potential wells.

The experimental setup used here to create Bose-Einstein condensates in the three-dimensional lattice potential (see methods section) is similar to our previous work[11,14,15]. Briefly, we start with a quasi-pure Bose-Einstein condensate of up to $2\times10^5$ $^{87}$Rb atoms in the $|F=2, m_F=2\rangle$ state in a harmonic magnetic trapping potential with isotropic trapping frequencies of ω=2π×24 Hz.

In order to transfer the magnetically trapped atoms into the optical lattice potential, we slowly increase the intensity of the lattice laser beams over a time of 80 ms so that a lattice potential depth $V_A$ of up to 11 $E_r$ is reached[11]. This potential depth $V_A$ is chosen so that the system is still completely in the superfluid regime[16]. We then rapidly increase the lattice potential depth to a value $V_B$ of up to 35 $E_r$ within a time of 50 μs so that the tunnel coupling between neighbouring potential wells becomes negligible. The timescale for the jump in the potential depth is chosen such that it is fast compared to the tunnelling time between neighbouring potential wells, but slow such that all atoms remain in the vibrational ground state of each well. Thereby we preserve the atom number distribution of the potential depth $V_A$ at the potential depth $V_B$.

We follow the dynamical evolution of the matter wave field after jumping to the potential depth $V_B$ by holding the atoms in the optical lattice for different times $t$. After these hold times we suddenly turn off the confining optical and magnetic trapping potentials and observe the resulting multiple matter wave interference pattern after a time-of-flight period of 16 ms. An example of such an evolution can be seen in Fig. 2, which shows the collapse and revival of the interference pattern over a time of 550 μs. This collapse and revival of the interference pattern is directly related to the collapses and revivals of the individual coherent matter wave fields in each potential well. It is important to note a crucial difference between the outcome of a collapse and revival experiment in a double well system and our multiple well system. In a double well system a perfect interference pattern would be observed in each single realization of the experiment for all times. However, when the matter wave fields have collapsed in both wells, this interference pattern would alternate randomly for each realization. Averaging over several single realizations would then yield the ensemble average value $\psi=0$ that indicates the randomness of the interference pattern associated with the collapse of the matter wave fields. For the multiple well setup used here, however, the interference pattern in a single realization of the experiment can only be observed if the matter wave fields in each potential well have constant relative phase to each other, which requires that $\psi \neq 0$. The matter wave field $\psi$ is therefore directly connected to the visibility of the multiple matter wave interference pattern in a single realization of the experiment.

In order to analyze the time evolution of the interference pattern quantitatively, we evaluate the number of atoms in the first and central order interference peaks $N_{coh}$ vs. the total number of atoms $N_{tot}$ in the time-of-flight images. In the optical lattice the matter wave field in each potential well $\psi_i(t) = \langle \phi_i(t) | \hat{a}_i | \phi_i(t) \rangle$ collapses and revives due to the nonlinear dynamics discussed above. In order to relate the time evolution of the global fraction of coherent atoms $N_{coh}/N_{tot}$ to such a single site time evolution $\psi_i(t)$ with $\bar{n}_i$ atoms on average on this lattice site, we sum the coherent fraction in each well over all $M$ lattice sites: $N_{coh}/N_{tot} = 1/N_{tot} \sum_{i=1}^{M} |\psi_i(t)|^2$. This sum can be converted into an integral using the classical probability distribution $W(\bar{n})$, which describes the probability of finding a lattice site with an average number of $\bar{n}$ atoms. If the single site dynamics is given by $\psi(t, \bar{n}, (U/J)_A, U_B)$, then the total number of coherent atoms can be determined by $N_{coh} = \int W(\bar{n}) |\psi(t, \bar{n}, (U/J)_A, U_B)|^2 d\bar{n}$. Using the Bose-Hubbard model and assuming a homogenous system, we are able to numerically calculate the initial atom number statistics on a single lattice site for finite $U/J$ up to $U/J \approx 20$ and small $\bar{n}$ using a Gutzwiller ansatz[13,17]. This allows us to predict the dynamical evolution of the matter wave field on a single lattice site $\psi(t, \bar{n}, (U/J)_A, U_B)$. Figure 3 shows the experimentally determined evolution of $N_{coh}/N_{tot}$ over time after jumping to the potential depth $V_B$. Up to five revivals are visible after which a damping of the signal prevents further detection of revivals.

The revival of the matter wave field in each potential well is expected to occur at multiple times of h/$U$, independent of the atom number statistics in each well. Therefore, in our inhomogeneous system, the macroscopic interference pattern should revive at the same times on all sites. Since the onsite matrix element $U$ increases for higher potential depths, we expect the revival time to decrease as $V_B$ increases. This is shown in Figure 4, where we have measured the revival period for different final potential depths $V_B$. We find excellent agreement between an ab initio calculation of h/$U$ from a band structure calculation and our data points. The revivals also directly prove the quantization of the underlying Bose field and provide a first experimental proof that collisions between atoms lead to a fully coherent collisional phase $U n(n-1)t/2\hbar$ of the $|n\rangle$-particle state over time even on the level of individual pairs of atoms.

As we increase our initial lattice potential depth $V_A$, we expect the atom number distribution in each well to become pronounced sub-Poissonian due to the increasing importance of the interactions as $U/J$ increases. This in turn should lead to an increase of the collapse time, which depends on the variance of the superimposed number states. We have verified this by measuring the collapse time for different initial potential depths $V_A$ (see Fig. 5a). One can clearly observe a significant increase in the collapse time, when jumping from higher potential depths. For example, when jumping from a potential depth of $V_A$=11 E$_r$, $t_c/t_{rev}$ is more than 50% larger than when jumping from an initial potential depth of $V_A$=4 E$_r$. This indicates that indeed the atom number distribution in each potential well has become sub-Poissonian since for our experimental parameters the average atom number per lattice site $\bar{n}_i$ remains almost constant when the lattice potential depth $V_A$ is increased. A comparison of the collapse time for different initial potential depths $V_A$ to a theoretical prediction is shown in Fig. 5b.

The observed collapse and revival of the macroscopic matter wave field of a Bose-Einstein condensate directly demonstrates a striking behaviour of ultracold matter beyond mean-field theories. Furthermore, the collapse times can serve as an independent efficient probe of the atom number statistics in each potential well.
In a next step it will be intriguing to start from a Mott insulating state and use the coherent collisions between single atoms, which have been demonstrated here, to create a many-atom entangled state[18-20]. This highly entangled state could then serve as a promising new starting point for quantum computing with neutral atoms[19,21].

**Methods**

**Optical lattices**

A three dimensional array of microscopic potential wells is created by overlapping three orthogonal optical standing waves at the position of the Bose-Einstein condensate. In our case the atoms are then trapped in the intensity maxima of the standing wave light field due to the resulting dipole force. The laser beams for the periodic potential are operated at a wavelength of λ=838 nm with beam waists of approx. 125 μm at the position of the Bose-Einstein condensate. This gaussian laser beam profile leads to an additional isotropic harmonic confinement of the atoms with trapping frequencies of 60 Hz for lattice potential depths of 20 $E_r$. Here $E_r$ denotes the recoil energy $E_r=\hbar^2 k^2/2m$, with $k=2\pi/\lambda$ being the wave vector of the laser light and *m* the mass of a single atom. In this configuration, we populate almost 150000 lattice sites with an average atom number per lattice site of up to 2.5 in the centre of the lattice. The lattice structure is of simple cubic type, with a lattice spacing of λ/2 and oscillation frequencies in each lattice potential well of approx. 30 kHz for a potential depth of 20 $E_r$.

We thank W. Zwerger, T. Esslinger, A. Görlitz, H. Briegel, E. Wright and I. Cirac for discussions and A. Altmeyer for help in the final stages of the experiment. This work was supported by the DFG.

Correspondence and requests for materials should be addressed to I.B.
(e-mail: imb@mpq.mpg.de).

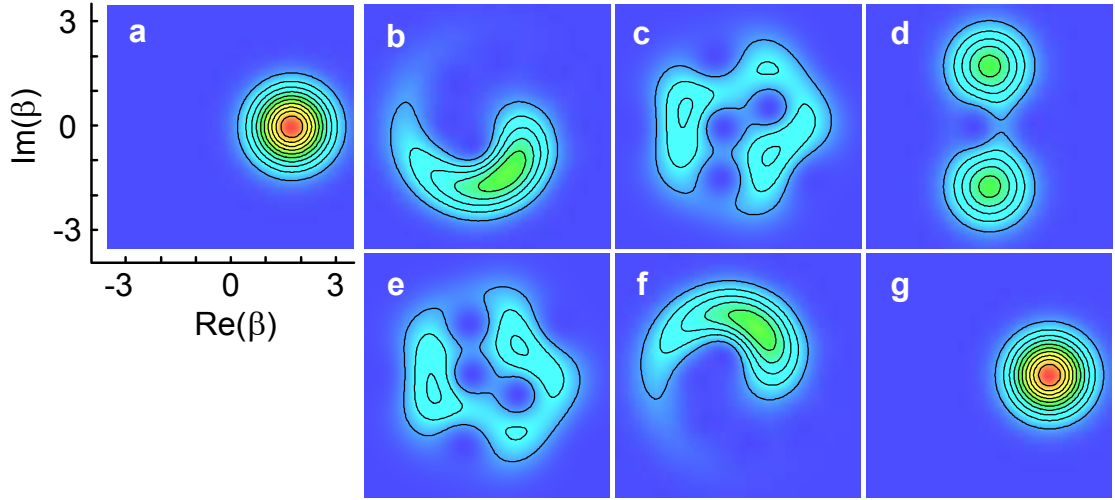

**Fig. 1.** Quantum dynamics of a coherent state due to cold collisions. The images **a-g** show the overlap $|\langle\beta|\alpha(t)\rangle|^2$ of an arbitrary coherent state $|\beta\rangle$ with complex amplitude β with the dynamically evolved quantum state $|\alpha\rangle(t)$ (see eq. 2) for an average number of $|\alpha|^2=3$ atoms at different times $t$. **a** $t=0$ h/$U$, **b** 0.1 h/$U$, **c** 0.4 h/$U$, **d** 0.5 h/$U$, **e** 0.6 h/$U$, **f** 0.9 h/$U$, and **g** h/$U$. Initially, the phase of the macroscopic matter wave field becomes more and more uncertain as time evolves **b**, but quite amazingly at $t_{rev}$/2 **d**, when the macroscopic field has collapsed such that $\psi \approx 0$, the system has evolved into an exact Schrödinger cat state of two coherent states. These two states are 180° out of phase and therefore lead to a vanishing macroscopic field $\psi$ at these times. More generally one can show that at certain rational fractions of the revival time $t_{rev}$, the system evolves into other exact superposition of coherent states, e.g. at $t_{rev}$/4 four coherent states or $t_{rev}$/3 three coherent states[2,4]. A full revival of the initial coherent state is then reached at $t$=h/$U$.

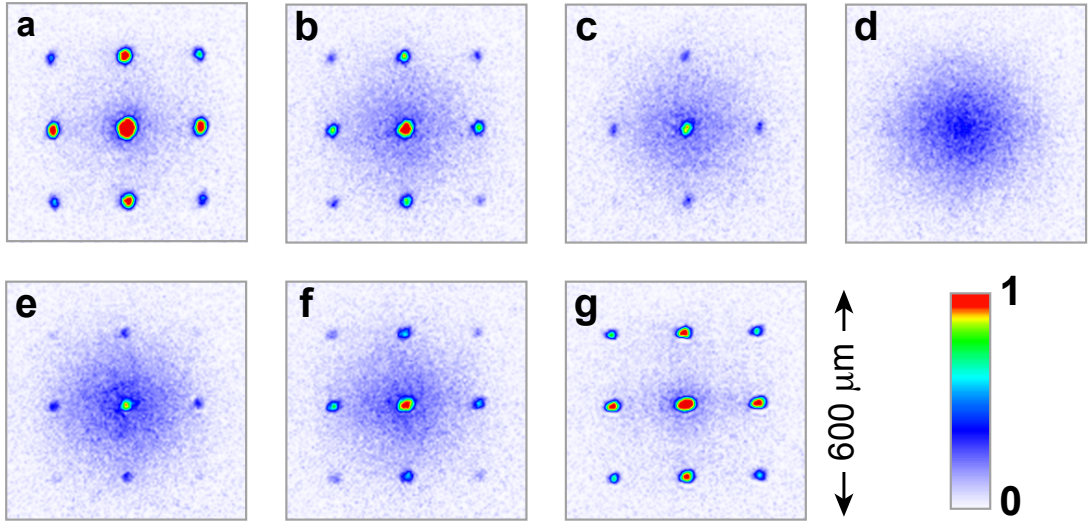

**Fig. 2** Dynamical evolution of the multiple matter wave interference pattern observed after jumping from a potential depth $V_A$=8 $E_r$ to a potential depth $V_B$=22 $E_r$ and a subsequent variable hold time $t$. After this hold time all trapping potentials were shut off and absorption images were taken after a time-of-flight period of 16 ms. The hold times $t$ were **a** 0 μs, **b** 100 μs, **c** 150 μs, **d** 250 μs, **e** 350 μs, **f** 400 μs, and **g** 550 μs. At first, a distinct interference pattern is visible, showing that initially the system can be described by a macroscopic matter wave with phase coherence between individual potential wells. Then after a time of approx. 250 μs the interference pattern is completely lost. The vanishing of the interference pattern is caused by a collapse of the macroscopic matter wave field in each lattice potential well. Remarkably, however, after a total hold time of 550 μs the interference pattern is almost perfectly restored, showing that the macroscopic matter wave field has revived. The atom number statistics in each well, however, remains constant throughout the dynamical evolution time. This is fundamentally different from the vanishing of the interference pattern with no further dynamical evolution, which is observed in the quantum phase transition to a Mott insulator, where Fock states are formed in each potential well. From the above images the number of coherent atoms $N_{coh}$ is determined by first fitting a broad two-dimensional Gaussian function to the incoherent background of atoms. The fitting region for the incoherent atoms excludes 130 μm × 130 μm squares around the interference peaks. Then the number of atoms in these squares is counted by a pixel-sum from which the number of atoms in the incoherent Gaussian background in these fields is subtracted to yield $N_{coh}$.

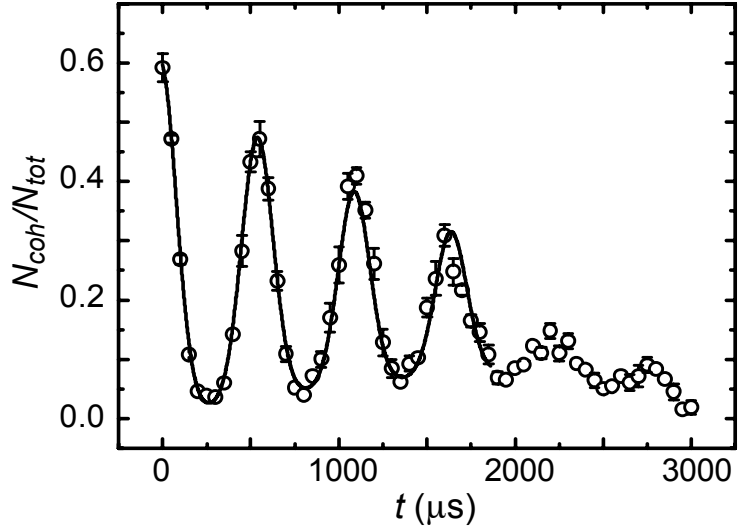

**Fig. 3** Number of coherent atoms relative to the total number of atoms monitored over time for the same experimental sequence as in Fig. 2. The solid line is a fit to the data assuming a sum of gaussians with constant widths and constant time separations including an exponential damping and a linear background. The damping is mainly due to the following reason: after jumping to a potential depth $V_B$ and thereby changing the external confinement and the onsite matrix element $U$ abruptly, we obtain a parabolic profile of the chemical potential over the cloud of atoms in the optical lattice, which leads to a broadening of the interference peaks over time. When the interference peaks become broader than the rectangular area in which they are counted, we cannot determine $N_{coh}$ correctly anymore, which explains the rather abrupt damping that can be seen e.g. between the third and fourth revival in the above figure. Furthermore, the difference in the on-site interaction matrix element $U$ of approx. 3% over the cloud of atoms contributes to the damping of $N_{coh}/N_{tot}$ over time. The finite contrast in $N_{coh}/N_{tot}$ of initially 60% can be attributed to atoms in higher order momentum peaks (approx. 10% of the total atom number), $s$-wave scattering spheres created during the expansion[14], a quantum depletion of the condensate for the initial potential depth of $V_A = 8\ E_r$ and a finite condensate fraction due to the finite temperature of the system.

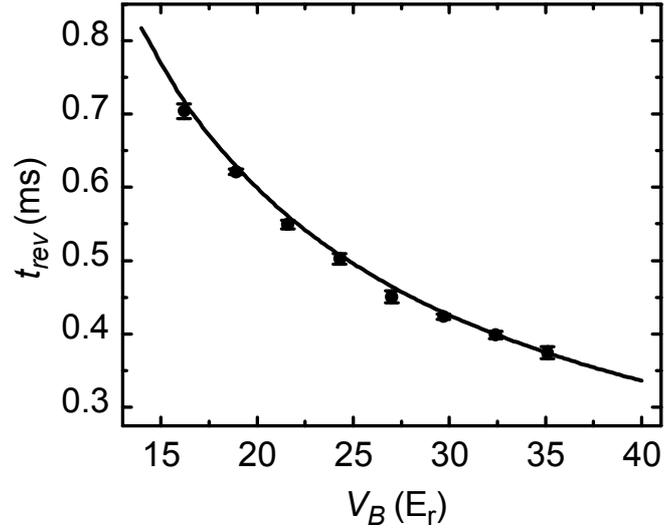

**Fig. 4** Revival period in the dynamical evolution of the interference pattern after jumping to different potential depths $V_B$ from a potential depth of $V_A$=5.5 $E_r$. The solid line is an ab-initio calculation of h/$U$ with no adjustable parameters based on a band structure calculation. In addition to the statistical uncertainties shown in the revival times, the experimental data points have a systematic uncertainty of 15% in the values for the potential depth.

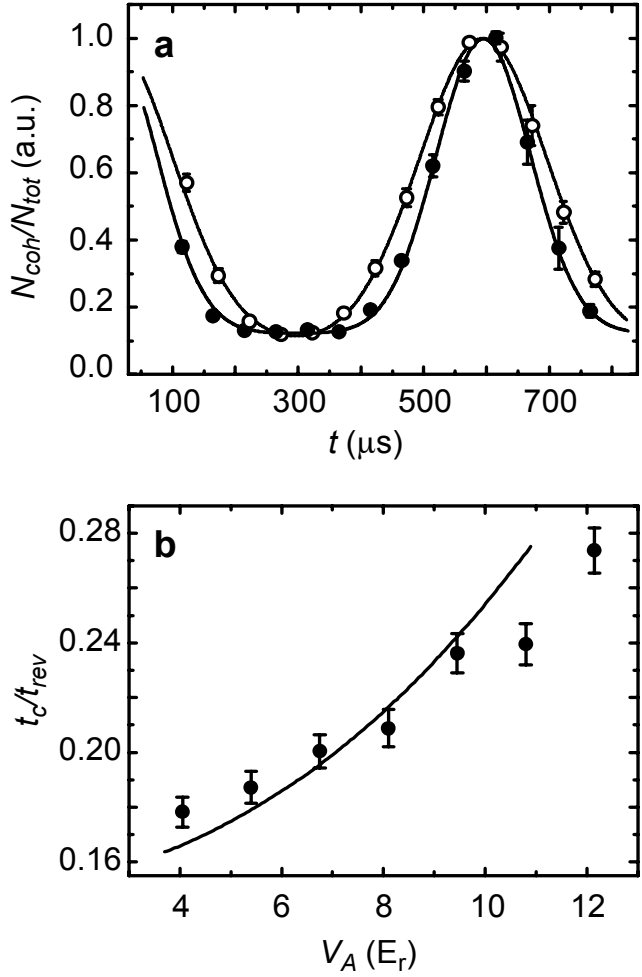

**Fig. 5** Influence of the atom number statistics on the collapse time. **a** First revival observed in the ratio $N_{coh}/N_{tot}$ after jumping from different initial potential depths $V_A$=4 $E_r$ (solid circles) and $V_A$=11 $E_r$ (open circles) to a potential depth of $V_B$= 20 $E_r$. The data have been scaled to the same height in order to compare the widths of the collapse times, where the contrast of the curve at $V_A$=11 $E_r$ was 20% smaller than that for $V_A$=4 $E_r$. The solid and dashed line are fits to the data assuming a sum of two Gaussians with constant widths $t_c$ (measured as the 1/e half width of the Gaussian), spaced by the corresponding revival time $t_r$ for the potential depth $V_B$=20 $E_r$. **b** Collapse time $t_c$ relative to the revival time $t_{rev}$ after jumping from different potential depths $V_A$ to a potential depth $V_B$= 20 $E_r$. The solid line is an ab-initio theoretical prediction based on the averaged time-evolution of the matter wave fields in each lattice potential well described in the text. Considering the systematic experimental uncertainties in the determination of the potential depths $V_A$ of about 15% and an uncertainty in the total atom number of about 20%, we find a reasonable agreement between both the experimental data and the theoretical prediction.